\documentclass[twocolumn,preprintnumbers,showkeys,superscriptaddress,nofootinbib]{revtex4}

\usepackage[dvips]{graphicx,color}
\usepackage{epsfig}
\usepackage{amsmath}
\usepackage{amssymb}
\usepackage{booktabs}
\usepackage[dvipdfm,bookmarks=true,bookmarksnumbered=true,bookmarkstype=toc]{hyperref} 

\allowdisplaybreaks[4]
\newcommand{\lsim}{\raise0.3ex\hbox{$\;<$\kern-0.75em\raise-1.1ex\hbox{$\sim\;$}}}
\newcommand{\gsim}{\raise0.3ex\hbox{$\;>$\kern-0.75em\raise-1.1ex\hbox{$\sim\;$}}}

\begin{document}

\title{Constraints from Color and/or Charge Breaking Minima in the $\nu$SSM}

\author{Tatsuo Kobayashi}
\email{kobayash@gauge.scphys.kyoto-u.ac.jp}
\affiliation{Department of Physics, Kyoto University, Kyoto 606-8502, Japan}

\author{Takashi Shimomura}
\email{stakashi@yukawa.kyoto-u.ac.jp}
\affiliation{Yukawa Institute for Theoretical Physics, Kyoto University, Kyoto 606-8502, Japan}

\date{\today}

\preprint{YITP-10-41}
\preprint{KUNS-2268}

\begin{abstract}
We consider a model where right-handed neutrinos and sneutrinos are introduced to the minimal supersymmetric standard model. 
In the scalar potential of this model, there exist trilinear and quartic terms in scalar potential that are proportional to Yukawa couplings of neutrinos. 
Due to these trilinear and quartic terms, Color and/or Charge Breaking (CCB) and Unbounded-From-Below (UFB) directions appear 
along which sneutrinos have a vacuum expectation value, making the vacuum of the electroweak symmetry breaking unstable. 
We analyze scalar potential of this model and derive necessary conditions for CCB and UFB directions to vanish.
\end{abstract}

\maketitle

\section{Introduction}\label{sec:introduction}
Neutrino oscillation experiments \cite{Fukuda:2002pe,Ahmad:2002jz,Ahn:2002up,Araki:2004mb,Ashie:2005ik} have confirmed that  
neutrinos have very tiny but non-zero masses. This is a clear evidence of physics beyond the standard model (SM) because 
neutrinos are massless in the SM. The simplest way to generate their tiny masses is to introduce right-handed neutrinos. 
There are two scenarios in this regard. One is that right-handed neutrinos are Majorana particles and neutrinos acquire 
masses via the famous seesaw mechanism \cite{Minkowski:1977sc,Yanagida:1979,Gell-mann:1979,Mohapatra:1979ia,Schechter:1980gr}. 
The other scenario is that right-handed neutrinos are Dirac particles and neutrinos obtain masses via electroweak symmetry 
breaking (EWSB). 

Combined these right-handed neutrino scenarios with a supersymmetric standard model, which we call $\nu$SSM, many works have been 
done so far. In the seesaw mechanism, it has been investigated recently that Majorana masses are as low as between 
$100$ GeV and $10$ TeV. In fact, such low scale Majorana masses can be realized as a consequence of supersymmetry (SUSY) breaking 
\cite{ArkaniHamed:2000bq,Borzumati:2000mc,MarchRussell:2004uf}. This class of models predict relatively small Yukawa couplings 
of neutrinos compared to those of other fermions.
In scenarios of Dirac neutrinos, tiny neutrino masses are solely explained by tiny Yukawa couplings. One might think it is 
unnatural because the Yukawa couplings of neutrinos are too small compared with those of other fermions. However, as was emphasized in 
\cite{Asaka:2005cn}, it is natural in 'tHooft's sense \cite{thooft1980} that a symmetry (i.e. chiral symmetry in the neutrino sector) is 
recovered in the limit of vanishing neutrino Yukawa coupling constants. In these scenarios, right-handed neutrinos and sneutrinos are light,  and 
therefore the scenarios are testable in astrophysical observations and terrestrial experiments. Studies of these scenarios are, e.g. 
the dark matter physics \cite{Asaka:2005cn,McDonald:2006if,Asaka:2006fs,Ishiwata:2009gs}, lepton flavour violation searches \cite{Ilakovac:2009jf} and 
collider physics \cite{Porod:2002zy,Cho:2006sm}.

The presence of the scalar partners generally leads color and/or charge breaking (CCB) directions and unbounded from below (UFB) 
directions \cite{Frere:1983ag,AlvarezGaume:1983gj,Derendinger:1983bz,Kounnas:1983td,Claudson:1983et,Drees:1985ie,Gunion:1987qv,
Komatsu:1988mt,Gamberini:1989jw}. 
Along CCB directions, the scalar potential has minima on which color and/or charge symmetries are spontaneously broken. 
The CCB minimum can be deeper than that of EWSB when the Yukawa coupling of the particle along CCB direction is 
small. Along UFB directions, the scalar potential has no global minimum and falls down to negative infinity. These directions make 
the vacuum of EWSB unstable, hence must be avoided. In the minimal supersymmetric extension of the SM (MSSM), conditions to avoid 
the UFB and CCB directions were systematically investigated in 
\cite{Casas:1995pd}\footnote{See also recent work \cite{Gioutsos:2002pd}}. 
Those conditions constrain soft SUSY breaking parameters, mainly trilinear couplings, 
and exclude a certain region of the parameter space of the MSSM. In the $\nu$SSM, due to right-handed sneutrinos, not only new UFB and CCB 
directions but also false EWSB directions appear. Along false EWSB directions, neither color nor charge symmetry is 
broken but Higgses and sneutrinos acquire large vacuum expectation values. Such minima result in too heavy masses of gauge bosons and are 
excluded by precise electroweak measurements. Since the EWSB vacuum can become unstable along these directions due to small neutrino Yukawa 
couplings, conditions to avoid those directions should be investigated. In this article, we refer false EWSB directions as CCB directions 
in view point of incorrect vacuum.

In this article, we consider the $\nu$SSM where either right-handed Dirac or Majorana (s)neutrinos are introduced to a supersymmetric standard model.
We assume that the Majorana masses are below or around TeV scale so that the neutrino Yukawa coupling is small as it is in the Dirac
neutrinos case. Then, we analyze the potential along UFB and CCB directions at tree level and derive necessary conditions to avoid 
dangerous minima and directions. Necessary conditions are generally modified due to radiative corrections \cite{Ford:1992mv,Casas:1994us}. 
The conditions from tree-level analysis coincide with 
those from one-loop analysis when analysis is performed at a scale that vacuum expectation values of Higgses with and without radiative 
corrections coincide  \cite{Ford:1992mv,Casas:1994us}. We assume that our analysis is performed at this scale. 

The outline of this article is organized as follows. In section \ref{sec:gener-prop-ufb-ccb}, we briefly review general properties of UFB
and CCB directions in the MSSM. Then we analyze the scalar potential and derive necessary conditions in Dirac and Majorana neutrino cases in
the section \ref{sec:constraints-from-ccb} and \ref{sec:constraints-from-ufb}. We show numerical results of constraints on the soft SUSY
breaking parameters in section \ref{sec:numerical-analysis}. Finally we summarize and discuss our analysis in the section
\ref{sec:summary-discussion}. The scalar potential of the MSSM and notations of fields and couplings are given in
Appendix~\ref{apd:scal-potent-mssm}, and EWSB of the MSSM is summarized in Appendix~\ref{sec:higgs-potent-electr}.  In
Appendix~\ref{apx:scal-potent-numssm}, $F$ terms and soft SUSY breaking terms in the $\nu$SSM are shown.

\section{General Properties for UFB directions and CCB minima in the MSSM}\label{sec:gener-prop-ufb-ccb}
We start our discussion with briefly reviewing general properties of UFB and CCB directions in the MSSM \cite{Casas:1995pd}. 
Following the general properties, it is possible to classify all dangerous directions 
in a field space. As was studied in \cite{Casas:1995pd}, there are three types of UFB directions and CCB directions, respectively. 
Throughout the main part of this paper, we refer $H_1$ and $H_2$ to a neutral component of down-type and up-type Higgs 
scalars, and use a symbol ``tilde'' to denote scalar partners of the SM fermions. 
Notations of couplings and fields and the scalar potential of the MSSM are summarized in Appendix.~\ref{apd:scal-potent-mssm}.

\subsection{General Properties of UFB directions}\label{sec:gener-prop-ufb}
In general, UFB directions appear along field configurations such that terms of scalars in a potential are vanishing or kept 
under control. Along these directions, the potential is unstable and its minimum is driven to negative infinity if quadratic terms of 
the fields are negative. Two general properties for UFB directions are shown below. 

{\it Property 1}. Trilinear scalar terms can not play a significant role along a UFB direction. This can be understood as follows. 
If a trilinear term does not vanish, $F$ terms give rise to (positive) quartic terms which lift the potential 
up for large values of scalar fields. Let us show an example. Suppose that the trilinear term corresponding to the Yukawa couplings 
of charged sleptons is non-vanishing, at least one of $F$ term of the scalar fields involved in the trilinear term is nonzero, e.g.
\begin{align}
 F_{\tilde{e}_R} = Y_e (H_1 \tilde{e}_L - H_1^- \tilde{\nu}_L),
\end{align}
where $H_1^-$ is a charged component of the down-type Higgs.
It is obvious that a positive quartic $F$ term which is proportional to $|Y_e|^2$ arises from the square of this term in the potential.

{\it Property 2}. Any UFB direction must involve $H_2$ and perhaps $H_1$. This is because the terms $|H_2|^2$ and $H_1 H_2$ can have 
negative soft masses for EWSB to successfully occur, while the other masses must be positive. Furthermore, since these  
terms are quadratic, all quartic terms coming from $F$ and $D$ terms must be vanishing or kept under control. Thus some additional 
fields are required except for $H_2$.

According to these properties, UFB directions are classified into three. A direction along which $H_1$ and $H_2$ have an equal vacuum
expectation value (vev) and other fields have no vev's, that is the so-called UFB-1 direction. Another direction, the so-called UFB-2 direction, 
is the direction with nonzero vev's of $H_1$, $H_2$ and $\tilde{L}$. Along the last direction called the UFB-3, $H_2$, $\tilde{L}$ and 
$\tilde{d}_L$, $\tilde{d}_R$ are non-vanishing. In the following, we show details of the UFB-2 and 3 directions. We will see that absence 
of the neutrino Yukawa coupling plays an essential role on these directions. 

Along the UFB-2 direction, left-handed sleptons have non-vanishing vev's to cancel quartic terms from $D$ terms. According to the property 1, 
the trilinear term involving left-handed sleptons must be vanishing in order not to give a quartic term proportional to the Yukawa
coupling squared. The only possibility for this direction is that the left-handed slepton has a vev along sneutrino direction 
since neutrinos are massless hence do not have Yukawa couplings. Then, the potential is given 
\begin{align}
 V_{\mathrm{UFB-2}} &= m_1^2 |H_1|^2 + m_2^2 |H_2|^2 -2 |m_3^2| |H_1| |H_2| \nonumber \\
  &+ m_{\tilde{L}}^2 |\tilde{L}|^2 + \frac{g_1^2 + g_2^2}{8}( |H_2|^2 - |H_1|^2 - |\tilde{L}|^2)^2. \label{eq:1}
\end{align}
The potential along the UFB-2 direction is obtained by minimizing Eq.~(\ref{eq:1}) with respect to $|\tilde{L}|$ and $|H_1|$,
\begin{align}
 V_{\mathrm{UFB-2}} = \left( m_2^2 + m_{\tilde{L}}^2 - \frac{|m_3^2|^2}{| m_1^2 - m_{\tilde{L}}^2 |} \right) |H_2|^2
  - \frac{2 m_{\tilde{L}}^4}{g_1^2 + g_2^2},\label{eq:2}
\end{align}
where 
\begin{align}
 &(m_1^2 - m_{\tilde{L}}^2)^2 > |m_3^2|^2,\\
 &|H_2|^2 > \frac{4 m_{\tilde{L}}^2}{(g_1^2 + g_2^2)(1-|m_3^2|^2/(m_1^2 - m_{\tilde{L}}^2)^2)},
\end{align}
are assumed. Notice that the condition for the minimum with respect to $|H_2|$, $\partial V/\partial |H_2| = 0$, 
can not be satisfied simultaneously, therefore $|H_2|$ is a free parameter in Eq.~(\ref{eq:2}).
The potential becomes unbounded from below if the quadratic term of $|H_2|$ is negative. Therefore the condition 
to avoid the UFB-2 direction  is 
\begin{align}
 m_2^2 + m_{\tilde{L}}^2 - \frac{|m_3^2|^2}{| m_1^2 - m_{\tilde{L}}^2 |} \ge 0.
\end{align}

Along the UFB-3 direction, $H_1$ is vanishing and vev's of down squarks are chosen to cancel $F$ term of $H_1$,
\begin{align}
 F_{H_1} = \mu H_2 + Y_d \tilde{d}_L \tilde{d}_R^\ast = 0.
\end{align}
Then, as we will explain in the next section, vev's of down squarks are much smaller than those of the Higgs and 
the sleptons, and can be neglected in the scalar potential.
Taking the vev's along $\tilde{d}_L = \tilde{d}_R^\ast = \tilde{d}$ so that $SU(3)$ $D$ term also vanishes, the potential 
becomes 
\begin{align}
 V_{\mathrm{UFB-3}} &= ( m_2^2 - |\mu|^2 ) |H_2|^2 + (m_{\tilde{Q}}^2 + m_{\tilde{d}_R}^2)|\tilde{d}|^2 + m_{\tilde{L}}^2 |\tilde{L}|^2 \nonumber \\
  &\quad + \frac{g_1^2 + g_2^2}{8}( |H_2|^2  + |\tilde{d}|^2 - |\tilde{L}|^2)^2, \label{eq:25}
\end{align}
where
\begin{align}
 |\tilde{d}|^2 = \frac{|\mu|}{|Y_d|}|H_2|.
\end{align}
Repeating the procedure of the UFB-2 direction, we can obtain the constraint preventing from the UFB-3 direction
\begin{align}
 m_2^2 - |\mu|^2 + m_{\tilde{L}}^2 \ge 0,
\end{align}
assuming
\begin{align}
 |H_2| > \sqrt{ \frac{|\mu|^2}{4 |Y_d|^2} + \frac{4 m_{\tilde{L}}^2}{g_1^2 + g_2^2} } - \frac{|\mu|}{2|Y_d|}.
\end{align}
It is important to emphasize here that the quartic terms from $F$ and $D$ terms can vanish simultaneously because 
the neutrino Yukawa coupling is absent.

\subsection{General Properties of CCB minima}
CCB minima appear along directions in which a negative trilinear term dominates a potential against quadratic and quartic terms at 
a certain region of field space. CCB minima become deeper as Yukawa couplings of scalars are smaller. In the following, 
we show five general properties of CCB minima in the MSSM.

{\it Property 1}. The deepest CCB direction involves only one particular trilinear soft term of one generation. When 
more than two trilinear terms are non-vanishing, quartic terms arising from $F$ terms are also non-vanishing. Different 
quartic terms hardly deepen the potential cooperatively, rather lift up the potential.

{\it Property 2}. It can not be determined a priori which trilinear coupling gives the strongest constraints. Non-vanishing trilinear 
terms lead quartic terms which are proportional to the square of a Yukawa coupling. Since the quartic terms are more important 
than the trilinear term for large values of fields, larger Yukawa couplings do no always deepen the potential.

{\it Property 3}. If the trilinear term under consideration has a very small Yukawa coupling, $D$ terms must be vanishing 
or negligible along the corresponding CCB direction. If $D$ terms are non-vanishing, it lifts up the potential faster than $F$ terms.
Then, that direction can not be the deepest direction.

{\it Property 4}. There are two directions to be explored for CCB. For example, for $A_u Y_u \tilde{Q} \cdot H_2 \tilde{u}_R^\ast$, 
one is the direction along which $H_2$, $\tilde{Q}$ and $\tilde{u}_R$ are non-vanishing, 
and $|\tilde{d}_L|^2 = |\tilde{d}_R|^2 = |\tilde{d}|^2$ so that $D_{SU(3)}$ and $F_{H_1}$ vanish. This direction is similar to UFB-3 
and called the direction (a) according to Casas, et al \cite{Casas:1995pd}. The other direction is along $H_1$, $H_2$ and 
$\tilde{Q}$, $\tilde{u}_R$ are nonzero. 
Possibly $\tilde{L}$ is also nonzero along this direction. The direction is similar to UFB-2 and called the direction (b). 

{\it Property 5}. There are two choices of the phases of soft SUSY breaking terms in the direction (b). For the same example as 
the above, the relevant soft terms are 
\begin{align}
 &2|A_u Y_u \tilde{Q} H_2 \tilde{u}_R| \cos\varphi_1 + 2 |\mu Y_u \tilde{Q} H_1 \tilde{u}_R| \cos\varphi_2 \nonumber \\
 &+ 2 |B\mu H_1 H_2|\cos\varphi_3,
\end{align}
where $\varphi_1$, $\varphi_2$ and $\varphi_3$ represent phases combined with signs of the couplings and phases of the fields.
If sign$(A_u) = -$ sign$(B)$, the three phases can be taken $\pi$, so the three terms are negative. On the other hand, 
if sign$(A_u) = $ sign$(B)$, two of them can be taken $\pi$ and the other one should be $0$. Therefore one of the three terms are positive. 
For the direction (a), only one term with an undetermined phase is $A_u Y_u \tilde{Q} H_2 \tilde{u}_R$. The sign of the term can always 
be taken negative by rotating the fields involved.

\section{Constraints from CCB minima with Dirac neutrinos}\label{sec:constraints-from-ccb}
In this section, we analyze the scalar potential of the $\nu$SSM with Dirac neutrinos. We consider not only the directions explained 
in the previous section (MSSM directions) but also new directions along which right-handed sneutrinos have vev's. 
In our analysis, we assume that only one sneutrino has non-vanishing vev.

\subsection{Constraints from MSSM UFB directions}
Let us consider the MSSM UFB-2 direction along which Higgses and left-handed sneutrinos have non-vanishing 
vev's. As we emphasized in the sec.~\ref{sec:gener-prop-ufb}, absence of the neutrino Yukawa coupling played an important role 
in the UFB-2 direction. The situation changes when the neutrino Yukawa coupling is introduced. $F$ terms given in 
Appendix~\ref{apx:scal-potent-numssm} can not be vanishing simultaneously. According to the property 2 of the UFB direction, a 
positive quartic term remains in the scalar potential, $V_{\mathrm{UFB-2}}^{\mathrm{Dirac}}$,
\begin{align}
  V_{\mathrm{UFB-2}}^{\mathrm{Dirac}} &= V_{\mathrm{UFB-2}} + |Y_\nu|^2 |H_2|^2 |\tilde{\nu}_L|^2,
\end{align}
where $V_{\mathrm{UFB-2}}$ is given in Eq.~(\ref{eq:2}). The last term lifts up the potential for large values  
of the fields. Thus, the MSSM UFB-2 direction disappears and turns to a CCB direction. We analyze this CCB direction below.

We parametrize the vev's for convenience,  
\begin{align}
 |\tilde{\nu}_L| = \alpha |H_2|,\quad |H_1| = \gamma |H_2|,
\end{align}
where $\alpha$ and $\gamma$ are real numbers. The potential is written using this parametrization,
\begin{align}
 V_{\mathrm{UFB-2}}^{\mathrm{Dirac}} = |Y_\nu|^2 F(\alpha,\gamma) \alpha^2 \gamma^2 |H_2|^4 + \hat{m}^2(\alpha,\gamma) |H_2|^2,\label{eq:19}
\end{align}
where
\begin{subequations}
\begin{align}
 F(\alpha,\gamma) &= \frac{1}{\gamma^2} + \frac{1}{\alpha^2 \gamma^2} f(\alpha,\gamma), \\
 f(\alpha,\gamma) &= \frac{1}{8} \frac{g_1^2 + g_2^2}{|Y_\nu|^2} (\alpha^2 + \gamma^2 -1)^2, \\
 \hat{m}^2(\alpha,\gamma) &= m_1^2 \gamma^2 -2|m_3^2| \gamma + m_2^2 + m_{\tilde{L}}^2 \alpha^2. \label{eq:23}
\end{align}
\end{subequations}
The minimum of the potential is obtained by differentiating Eq.~(\ref{eq:19}) with respect to $|H_2|$,
\begin{align}
 |H_2|^2_{ext} = -\frac{1}{2} \frac{\hat{m}^2(\alpha,\gamma)}{|Y_\nu|^2 F(\alpha,\gamma) \alpha^2 \gamma^2}, \label{eq:20}
\end{align}
where $|H_2|_{ext}$ is the vev of $H_2$ at extremal. Here we assumed that $\hat{m}^2(\alpha,\gamma)$ is negative. 
According to the property 3 of the CCB direction, we set $\alpha^2 = 1 - \gamma^2$ to cancel the $D$ term or $f(\alpha,\gamma)$. 
Inserting Eq.~(\ref{eq:20}) into the potential, the minimum is expressed 
\begin{align}
 V_{\mathrm{UFB-2~min}}^{\mathrm{Dirac}} = -\frac{1}{4} \frac{\big(\hat{m}^2(\gamma)\big)^2}{|Y_\nu|^2 (1 - \gamma^2)},
\end{align}
where 
\begin{align}
 \hat{m}^2(\gamma) = ( m_1^2 - m_{\tilde{L}}^2) \gamma^2 -2|m_3^2| \gamma + m_2^2 + m_{\tilde{L}}^2. \label{eq:24}
\end{align}
The minimum would be much deeper than that of the EWSB, (\ref{eq:18}), because the neutrino Yukawa coupling is very small. 
A necessary condition to avoid the dangerous minimum is that $\hat{m}^2$ is positive for any $\gamma$. It imposes a constraint 
on the soft masses as
\begin{align}
 0 \le |m_3^2|^2 - m_1^2 m_2^2 \le m_{\tilde{L}}^2 (m_1^2 - m_2^2 + m_{\tilde{L}}^2), \label{eq:21}
\end{align}
where the left inequality is imposed by Eq.~(\ref{eq:14}). The constraint forbids a small soft mass for the left-handed sleptons 
unless $|m_3^2|^2$ is close to $m_1^2 m_2^2$.

For the MSSM UFB-3 direction, the same quartic term remains in the potential,
\begin{align}
 V_{\mathrm{UFB-3}}^{\mathrm{Dirac}} &= V_{\mathrm{UFB-3}} + |Y_\nu|^2 |H_2|^2 |\tilde{\nu}_L|^2,
\end{align}
and alters the MSSM UFB-3 direction to a CCB direction. $V_{\mathrm{UFB-3}}$ is given in (\ref{eq:25}). As is shown below, 
the vev's of the Higgses and sneutrinos are of order $m_{soft}/Y_\nu$ where $m_{soft}$ is a typical scale of the soft SUSY breaking masses. 
These vev's are much larger than those of the down squarks. Therefore, we can neglect down squarks in the following discussion. 
Similarly to the UFB-2 direction, the potential is expressed using a parametrization, $|\tilde{\nu}_L| = \alpha |H_2|$,
\begin{align}
 V_{\mathrm{UFB-3}}^{\mathrm{Dirac}} = |Y_\nu|^2 F(\alpha) \alpha^2 |H_2|^4 + \hat{m}^2(\alpha) |H_2|^2,
\end{align}
where
\begin{subequations}
\begin{align}
 F(\alpha) &= 1 + \frac{1}{\alpha^2} f(\alpha), \\
 f(\alpha) &= \frac{1}{8}\frac{(g_1^2 + g_2^2)}{|Y_\nu|^2} (\alpha^2 -1)^2, \\
 \hat{m}^2(\alpha) &= m_2^2 - |\mu|^2 + m_{\tilde{L}}^2 \alpha^2.
\end{align}
\end{subequations}
Minimizing the potential with respect to $|H_2|$, the value of the $|H_2|$ at extremal, $|H_2|_{ext}$, is obtained,
\begin{align}
 |H_2|^2_{ext} &= -\frac{1}{2} \frac{\hat{m}^2(\alpha)}{|Y_\nu|^2 F(\alpha) \alpha^2},
\end{align}
and the minimum of the potential is given by
\begin{align}
 V_{\mathrm{UFB-3~min}}^{\mathrm{Dirac}} = -\frac{1}{4}\frac{(\hat{m}^2)^2}{|Y_\nu|^2},
\end{align}
where $\alpha^2 = 1$ is used and $\hat{m}^2 = \hat{m}^2(\alpha^2=1)$. Again, the minimum is much deeper than that of the 
EWSB, (\ref{eq:18}). A necessary condition to avoid the CCB minimum is 
\begin{align}
 m_2^2 - |\mu|^2 + m_{\tilde{L}}^2 \ge 0. \label{eq:22}
\end{align}

\subsection{Constraint from CCB-1 minimum}
In the following, we analyze the scalar potential along CCB directions. Along the CCB directions of the MSSM, there are no important 
modifications on the constraints given in \cite{Casas:1995pd} since the trilinear term involving right-handed sneutrinos is vanishing 
and the quartic term proportional to the neutrino Yukawa coupling is very small. Once we consider directions that right-handed sneutrinos 
are non-vanishing, there appear new directions along which the minimum can become much deeper than that of EWSB. We focus our analysis 
on new CCB directions and derive constraints to evade such CCB minima. 

Firstly, we consider a direction similar to the MSSM CCB direction (a). From the properties 1 and 3 of the CCB direction, we assume 
\begin{align}
 &H_2,~\tilde{\nu}_L,~\tilde{\nu}_R \neq 0,\\
 &|\tilde{d}_L|^2 = |\tilde{d}_R|^2 = |d|^2,
\end{align}
and sign$(A_\nu) = -$sign$(B)$ for simplicity. Other fields are vanishing. The assumption $|\tilde{d}_L|^2 = |\tilde{d}_R|^2$ 
is made to cancel the $SU(3)~D$ term. Furthermore $\tilde{d}_L \tilde{d}_R^\ast$ is chosen to
cancel $F_{H_1}$. Analogous to the MSSM UFB-3 direction, the vev's of the Higgses and the sneutrinos 
are inversely proportional to the Yukawa coupling of neutrinos and are much larger than those of the down squarks. 
Hence we neglect down squarks in the potential. 

The scalar potential from $F$, $D$ terms and the soft SUSY breaking terms is given in Appendix~\ref{apx:scal-potent-numssm} and also 
Appendix~\ref{apd:scal-potent-mssm}. Following the procedure of the MSSM UFB-3 direction, we parametrize vev's as
\begin{align}
 |\tilde{\nu}_L| = \alpha |H_2|,~~ |\tilde{\nu}_R^\ast| = \beta |H_2|,
\end{align}
where $\alpha$ and $\beta$ are real numbers. Then, the scalar potential is written
\begin{align}
 V_{\mathrm{CCB-1}}^{\mathrm{Dirac}} &= |Y_\nu|^2 F(\alpha, \beta) \alpha^2 \beta^2 |H_2|^4 -2 |Y_\nu| \hat{A} \alpha \beta |H_2|^3 \nonumber \\
   &\quad + \hat{m}^2(\alpha,\beta) |H_2|^2, \label{eq:3}
\end{align}
where 
\begin{subequations}
\begin{align}
 F(\alpha,\beta) &= 1 + \frac{1}{\alpha^2} + \frac{1}{\beta^2} + \frac{1}{\alpha^2 \beta^2} f(\alpha), \\
 f(\alpha) &= \frac{1}{8} \frac{g_1^2 + g_2^2}{|Y_\nu|^2} (\alpha^2 -1)^2, \\
 \hat{A} &= |A_\nu|, \\
 \hat{m}^2(\alpha,\beta) &= m_{H_2}^2 + m_{\tilde{L}}^2 \alpha^2 + m_{\tilde{\nu}_R}^2 \beta^2.
\end{align}
\end{subequations}
Here, $|H_2|_{ext}$ is obtained by minimizing the right-hand side of Eq.~(\ref{eq:3}) with respect to $|H_2|$ for fixed values 
of $\alpha$ and $\beta$, 
\begin{align}
 |H_2|_{ext} &= \frac{3 \hat{A}}{4 |Y_\nu| F(\alpha,\beta) \alpha \beta} 
   \left( 1 + \sqrt{1 - \frac{8 \hat{m}^2(\alpha,\beta) F(\alpha,\beta)}{ 9 \hat{A}^2}} \right).\label{eq:4}
\end{align}
The minimum is given by inserting Eq.~(\ref{eq:4}) into Eq.~(\ref{eq:3}),
\begin{align}
 V_{\mathrm{CCB-1~min}}^{\mathrm{Dirac}}= -\frac{1}{2} \alpha \beta |H_2|_{ext}^2 
  \left(Y_\nu \hat{A} |H_2|_{ext} - \frac{\hat{m}^2(\alpha,\beta)}{\alpha \beta} \right). \label{eq:5}
\end{align}
The CCB-1 minimum would be much deeper than the EWSB minimum,~(\ref{eq:18}), because it is inversely proportional to $|Y_\nu|^2$. 
A necessary condition to avoid the minimum is that $V_{\mathrm{CCB-1~min}}$ becomes positive, which reads 
\begin{align}
  |A_\nu|^2 \le \frac{1 + 2 \beta^2}{\beta^2} \big( m_{H_2}^2 + m_{\tilde{L}}^2 + m_{\tilde{\nu}_R}^2 \beta^2 \big), \label{eq:6}
\end{align}
where $\alpha^2 = 1$ is set to cancel $D$ term, according to the property 3 of the CCB direction. 
We can further simplify the condition by minimizing the right-hand side of Eq.~(\ref{eq:6}). Differentiating the right-hand side with 
respect to $\beta^2$, $\beta^2$ for extremal is obtained,
\begin{align}
 \beta^4_{ext} = \frac{ m_{H_2}^2 + m_{\tilde{L}}^2 }{2 m_{\tilde{\nu}_R}^2}, \label{eq:7}
\end{align}
and inserting Eq.~(\ref{eq:7}), the condition becomes 
\begin{align}
  |A_\nu| &\le \sqrt{2 (m_{H_2}^2 + m_{\tilde{L}}^2) } + m_{\tilde{\nu}_R}, \label{eq:8}
\end{align}
and the trilinear term is bounded from above. It is important to notice that the condition (\ref{eq:22}) appears 
in the right-hand side. Therefore one can avoid both dangerous CCB minima once the constraint (\ref{eq:8}) is satisfied.

\subsection{Constraint from CCB-2 minimum}
Next, we analyze a direction similar to the MSSM CCB direction (b). We assume 
\begin{align}
 H_1,~ H_2,~\tilde{\nu}_L,~\tilde{\nu}_R \neq 0,
\end{align}
and other fields are zero. It is also assumed that sign$(A_\nu) = -$sign$(B)$.
Note that neither color nor charge symmetry is broken along this direction. Instead vev's of Higgses and sneutrinos are so large that 
weak gauge bosons are too heavy, therefore EWSB does not occur correctly. As we mentioned in the introduction, we call this direction as 
CCB direction. Then, we parametrize vev's as 
\begin{align}
 |\tilde{\nu}_L| = \alpha |H_2|,\quad |\tilde{\nu}_R| = \beta |H_2|,\quad |H_1| = \gamma |H_2|,
\end{align}
where $\alpha$, $\beta$ and $\gamma$ are real numbers. Then, the scalar potential is written
\begin{align}
 V_{\mathrm{CCB-2}}^{\mathrm{Dirac}} &= Y_\nu^2 F(\alpha, \beta, \gamma) \alpha^2 \beta^2 |H_2|^4
    -2 Y_\nu \hat{A}(\gamma) \alpha \beta |H_2|^3 \nonumber \\
   &\quad + \hat{m}^2(\alpha,\beta,\gamma) |H_2|^2,
\end{align}
where
\begin{subequations}
 \begin{align}
   F(\alpha,\beta,\gamma) &= 1 + \frac{1}{\alpha^2} + \frac{1}{\beta^2} + \frac{1}{\alpha^2 \beta^2} f(\alpha,\gamma), \\
 f(\alpha,\gamma) &= \frac{1}{8} \frac{g_1^2 + g_2^2}{|Y_\nu|^2} (\alpha^2 + \gamma^2 -1)^2, \\
 \hat{A}(\gamma) &= |A_\nu| + \gamma |\mu|, \\
 \hat{m}^2(\alpha,\beta,\gamma) &= m_1^2 \gamma^2 + m_2^2 + m_{\tilde{L}}^2 \alpha^2 + m_{\tilde{\nu}_R}^2 \beta^2 \nonumber \\
  &\quad -2 |m_3^2| \gamma,
 \end{align}
\end{subequations}
and sign$(A_\nu) = -$sign$B$ is assumed. The constraint from the CCB-2 direction is obtained by iterating the same procedure of the CCB-1
or the MSSM UFB-2 direction,
\begin{align}
  |A_\nu| \le - |\mu| \gamma +\left( 1 + \frac{2-\gamma^2}{1-\gamma^2} \beta^2_{ext}(\gamma) \right) m_{\tilde{\nu}_R}, \label{eq:9}
\end{align}
where $\beta_{ext}(\gamma)$ is 
\begin{align}
 \beta^4_{ext}(\gamma) = \frac{1-\gamma^2}{2-\gamma^2} 
   \frac{(m_1^2 - m_{\tilde{L}}^2) \gamma^2 -2 |m_3^2| \gamma + m_2^2 + m_{\tilde{L}}^2}{m_{\tilde{\nu}_R}^2}, \label{eq:10}
\end{align}
and $\alpha^2 = 1 - \gamma^2$ is used. It is seen that the constraint (\ref{eq:21}) is 
satisfied, hence the MSSM UFB-2 can be evaded if $\beta^4_{ext}(\gamma)$ is positive for any $\gamma$. 

The stringent constraint on $|A_\nu|$ is given by minimizing the right-hand side of Eq.~(\ref{eq:9}) with respect to $\gamma$, 
but it is not easy to obtain $\gamma_{ext}$ analytically because of complications. Therefore we just give an equation 
that $\gamma_{ext}$ must be satisfied,
\begin{align}
  &-|\mu|\beta^2_{ext}(\gamma_{ext}) + m_{\tilde{\nu}_R} \bigg[ \frac{\gamma_{ext}}{(1-\gamma^2_{ext})^2} \beta^4_{ext}(\gamma_{ext}) \nonumber \\ 
  &\quad + \frac{(m_1^2 - m_{\tilde{L}}^2) \gamma_{ext} - |m_3^2|}{m_{\tilde{\nu}_R}^2} 
 \bigg] =0. \label{eq:11} 
\end{align}
Equation (\ref{eq:11}) should be solved numerically. If $\gamma_{ext}$ is negative, $\gamma_{ext}=0$ is chosen and the condition from CCB-1
direction is obtained by replacing $m_{H_2}^2$ with $m_2^2$. If $\gamma_{ext}$ is larger than unity, $\gamma_{ext}=1$ and $\alpha = 0$ are 
chosen. Then, the potential becomes
\begin{align}
 V_{\mathrm{CCB-2}}^{\mathrm{Dirac}} &= Y_\nu^2 \beta^2 |H_2|^4 + \hat{m}^2(0,\beta,1) |H_2|^2,
\end{align}
where
\begin{align}
 \hat{m}^2(0,\beta,1) &= m_1^2 + m_2^2 -2 |m_3^2| + m_{\tilde{\nu}_R}^2 \beta \nonumber \\
                    &~ \ge m_{\tilde{\nu}_R}^2 > 0,
\end{align}
and we used Eq.~(\ref{eq:14}). Thus, the potential has a global minimum at 
$H_1 = H_2 = \tilde{\nu}_L = \tilde{\nu}_R = 0$.

\subsection{Constraint from CCB-3 minimum}
The CCB-3 direction is defined as the CCB-2 with sign$(A_\nu) =$ sign$(B)$. Along this direction, one of the signs among $|A_\nu|$, $|\mu|$ 
and $|m_3^2|$ is flipped according to the property 5.

When the sign of $|A_\nu|$ or $|\mu|$ is flipped, the condition to avoid the CCB minimum is given
\begin{align}
 |A_\nu| \le |\mu| \gamma + \left( 1 + \frac{2-\gamma^2}{1-\gamma^2} \beta^2_{ext}(\gamma) \right) m_{\tilde{\nu}_R},\label{eq:12}
\end{align}
where $\beta^4_{ext}$ is the same as Eq.~(\ref{eq:10}). 

When the sign of $|m_3^2|$ is flipped, the constraint becomes
\begin{align}
 |A_\nu|  \le |\mu| \gamma + \left(1 + \frac{2-\gamma^2}{1-\gamma^2}\tilde{\beta}^2_{ext}(\gamma) \right) m_{\tilde{\nu}_R},
\end{align}
where
\begin{align}
 \tilde{\beta}^4_{ext}(\gamma) = \frac{1-\gamma^2}{2-\gamma^2} 
  \frac{(m_1^2 - m_{\tilde{L}}^2) \gamma^2 + 2 |m_3^2| \gamma + m_2^2 + m_{\tilde{L}}^2}{m_{\tilde{\nu}_R}^2}.
\end{align}
The corresponding sign of $|\mu|$ or $|m_3^2|$ in Eq. (\ref{eq:11}) for $\gamma_{ext}$ should be also flipped appropriately.

\section{Constraints from UFB and CCB minima with Majorana neutrinos}\label{sec:constraints-from-ufb}
We consider the $\nu$SSM with Majorana neutrinos and analyze its potential given in Appendix~\ref{apx:scal-potent-numssm}. Differences 
from the Dirac case are the Majorana mass term in the superpotential and the corresponding soft SUSY breaking mass. These additional terms 
result in linear and quadratic terms of the right-handed sneutrinos in the scalar potential.

It is immediately understood that the constraints from the MSSM UFB directions are the same as those of the Dirac case, 
(\ref{eq:21}) and (\ref{eq:22}), because the right-handed sneutrinos do not have vev's. There appears a new UFB direction along
\begin{align}
 \tilde{\nu}_R \neq 0,\quad \mathrm{other~fields}=0,
\end{align}
and sign$(B_\nu M_R) = -1$. The potential along this direction is given
\begin{align}
  V_{\mathrm{UFB}}^{\mathrm{Majorana}}=( m_{\tilde{\nu}_R}^2 - |B_\nu M_R| + |M_R|^2 )|\tilde{\nu}_R|^2,
\end{align}
and it is unbounded from below unless 
\begin{align}
 m_{\tilde{\nu}_R}^2 - |B_\nu M_R| + |M_R|^2 \ge 0.
\end{align}

Along the CCB directions, we simply show results because the procedure to find the conditions is the same as in the Dirac case. 
The conditions are obtained by making replacements,
\begin{align}
 |A_\nu| &~\rightarrow~ |A_\nu| + |M_R|, \\
 m_{\tilde{\nu}_R} &~\rightarrow~ m_{\tilde{\nu}_R} + |B_\nu M_R| + |M_R|^2,
\end{align}
where sign$(B_\nu M_R) = 1$ is assumed. From the CCB-1 minimum, it is given from Eq.~(\ref{eq:8}),
\begin{align}
  |A_\nu| &\le - |M_R| + \sqrt{ 2(M_{H_2}^2 + M_{\tilde{L}}^2) } \nonumber \\
   &\quad + \sqrt{m_{\tilde{\nu}_R}^2 + |B_\nu M_R| + |M_R|^2}.
\end{align}
From the CCB-2 minimum, the condition is obtained from Eq.~(\ref{eq:9}),
\begin{align}
  |A_\nu| &\le -( |\mu|\gamma_{ext} + |M_R| ) + \left( 1 + \frac{2-\gamma^2}{1-\gamma^2} \beta^2_{ext}(\gamma_{ext}) \right) \nonumber \\
   &\qquad \times \sqrt{m_{\tilde{\nu}_R}^2 + |B_\nu M_R| + |M_R|^2},
\end{align}
where
\begin{align}
 \beta^4_{ext}(\gamma) = \frac{1-\gamma^2}{2-\gamma^2} 
  \frac{ (m_1^2 - M_{\tilde{L}}^2)\gamma^2 -2 |m_3^2|\gamma + m_2^2 + M_{\tilde{L}}^2}{m_{\tilde{\nu}_R}^2 + |B_\nu M_R| + |M_R|^2}.
\end{align}
Here $\gamma_{ext}$ is determined from
\begin{align}
   &-|\mu|\beta^2_{ext}(\gamma_{ext}) + m_{\tilde{\nu}_R} \bigg[ \frac{\gamma_{ext}}{(1-\gamma^2_{ext})^2} \beta^4_{ext}(\gamma_{ext}) \nonumber \\ 
  &\quad + \frac{(m_1^2 - m_{\tilde{L}}^2) \gamma_{ext} - |m_3^2|}{m_{\tilde{\nu}_R}^2 + |B_\nu M_R| + |M_R|^2} 
 \bigg] =0.
\end{align}
Along the CCB-3 direction, the same replacement should be done. For the case of sign$(B_\nu M_R)=-1$, the sign of $|B_\nu M_R|$ is 
flipped.

\section{Numerical Analysis}\label{sec:numerical-analysis}
We show numerical results for the Dirac neutrino case to demonstrate a strategy to constrain the soft SUSY parameters with the conditions
from UFB and CCB-1, 2, i.e.  (\ref{eq:21}), (\ref{eq:8}) and (\ref{eq:9}). The conditions are important for relatively light sneutrinos,
therefore we vary masses of sneutrinos fixing the Higgs masses. 

We calculate the Higgs soft masses using SPS1a point\cite{Allanach:2002nj} as an example. The parameters  
we use are
\begin{align}
 &\mu = 3.57\times 10^2,~B=47.2,\\
 &m_{H_1}^2=3.24 \times 10^4,~m_{H_2}^2 = -1.28 \times 10^5,
\end{align}
in the unit of GeV, and $m_{\tilde{L}}$ is taken as $360$ and $560$ GeV so that $\beta^4_{ext}$ along the CCB-1 direction 
is positive. It is assumed that sign$(A_\nu)= -$ sign$(B)$. The EWSB occurs correctly and the lighter Higgs mass is above $114$ GeV with these 
parameters. 

\begin{figure}[t]
\begin{tabular}{c}
 \includegraphics[width=80mm]{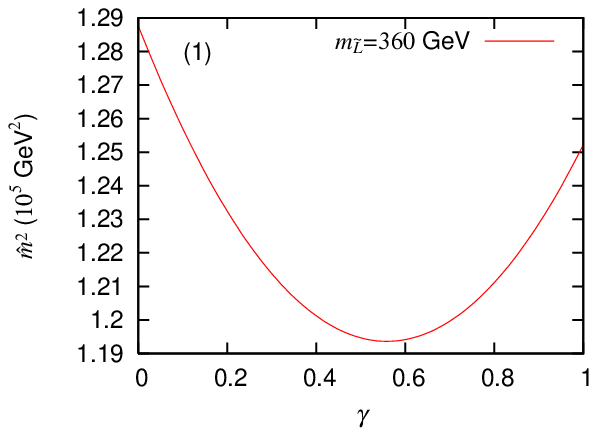} \\ 
 \includegraphics[width=80mm]{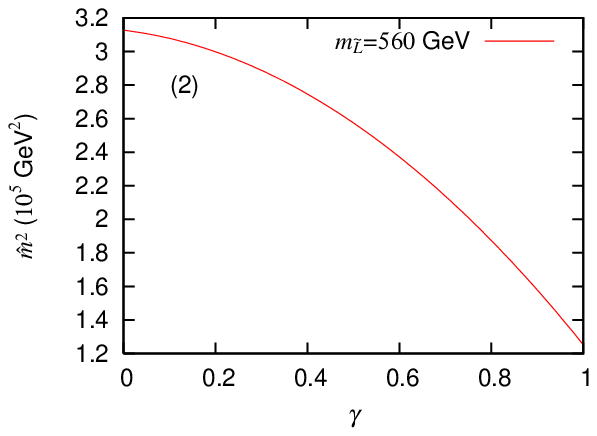} 
\end{tabular}
\caption{$\hat{m}^2(\gamma)$ in terms of $\gamma$. $m_{\tilde{L}}$ is taken $360$ GeV in the upper panel, 
and $560$ GeV in the lower panel, respectively.}
\label{fig:mhat}
\end{figure}
We start with checking that Eq.~(\ref{eq:24}) is positive between $0 \le \gamma \le 1$ for a given set of the parameters.
Figures \ref{fig:mhat} show Eq.~(\ref{eq:24}) with respect to $\gamma$. The upper panel is for $m_{\tilde{L}}= 360$ GeV and the lower 
panel is for $m_{\tilde{L}}=560$ GeV. It is seen that $\hat{m}^2$ is positive in both cases, and hence Eq.~(\ref{eq:21}) is satisfied. 
 
\begin{figure}[t]
\begin{tabular}{c}
 \includegraphics[width=80mm]{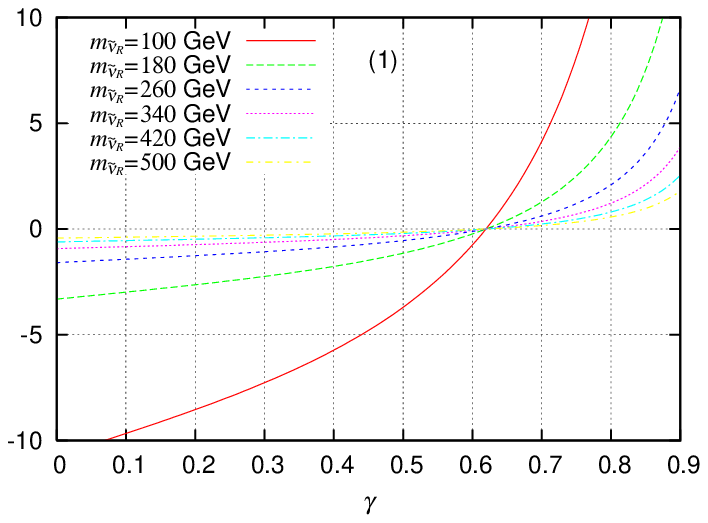} \\ 
 \includegraphics[width=80mm]{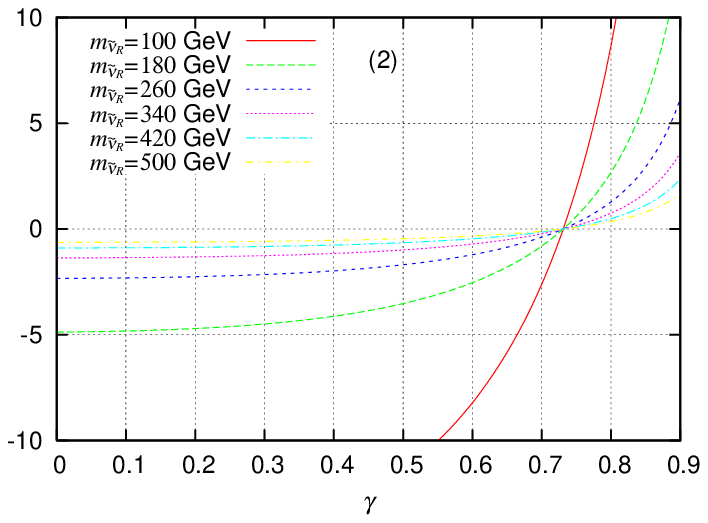} 
\end{tabular}
\caption{The left-hand side of the equation (\ref{eq:11}) normalized by $m_{\tilde{\nu}_R}$ in terms of $\gamma$ for 
various $m_{\tilde{\nu}_R}$. The values of $m_{\tilde{\nu}_R}$ are shown in the figures. The left-handed slepton soft mass 
is $360$ GeV in Fig.~\ref{fig:gamma_ext}.(1), and $560$ GeV in Fig.~\ref{fig:gamma_ext}.(2), respectively.}
\label{fig:gamma_ext}
\end{figure}
Secondly, we calculate $\gamma_{ext}$ using Eq.~(\ref{eq:11}). Figures \ref{fig:gamma_ext} show the left-hand 
side of Eq.~(\ref{eq:11}) normalized by $m_{\tilde{\nu}_R}$ in terms of $\gamma$. $m_{\tilde{\nu}_R}$ is varied from $100$ GeV to $500$
GeV. The mass of the left-handed slepton for each curve is shown in figures. $m_{\tilde{L}}$ is $360$ GeV in Fig.~\ref{fig:gamma_ext}.(1) 
and $560$ GeV in Fig.~\ref{fig:gamma_ext}.(2). The crossing point of each curve to zero corresponds to 
$\gamma_{ext}$. It is seen that $\gamma_{ext}$ is independent of $m_{\tilde{\nu}_R}$. This is because $m_{\tilde{\nu}_R}$ can be factored 
out by inserting the concrete form of $\beta_{ext}$. From figures, we can obtain $\gamma_{ext}=0.62$ and $0.73$ respectively. 
It is also seen that $\gamma_{ext}$ for $m_{\tilde{L}}=560$ GeV is larger than that for $360$ GeV. Generally $\gamma_{ext}$ becomes 
larger as $m_{\tilde{L}}$ increases for fixed values of other parameters although the dependence of $\gamma_{ext}$ 
on other parameters is so complicated that it can not be understood easily.

\begin{figure}[t]
\begin{tabular}{c}
 \includegraphics[width=80mm]{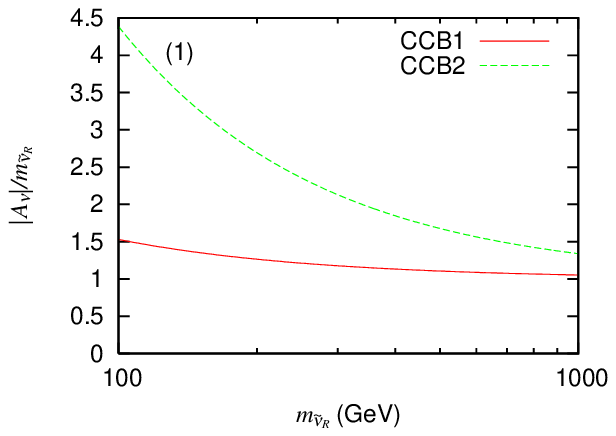} \\ 
 \includegraphics[width=80mm]{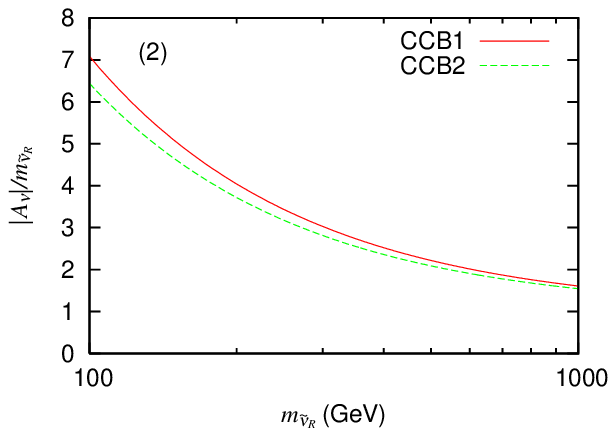} 
\end{tabular}
\caption{The constraints from CCB minimum normalized with $m_{\tilde{\nu}_R}$ in terms of $m_{\tilde{\nu}_R}$. 
The left-handed slepton soft mass is $360$ GeV 
in Fig.~\ref{fig:Anu_over_MR}.(1), and $560$ GeV in Fig.~\ref{fig:Anu_over_MR}.(2), respectively. The solid (red) curve represents Eq.~(\ref{eq:8}) and the dashed (green) represents Eq.~(\ref{eq:9}).}
\label{fig:Anu_over_MR}
\end{figure}
Thirdly, the constraints from CCB-1 and CCB-2 are calculated. In Figs.~\ref{fig:Anu_over_MR}, we plot the constraints normalized 
with $m_{\tilde{\nu}_R}$ by varying the right-handed slepton mass from $100$ to $1000$ GeV. The mass of the left-handed slepton is taken  
$360$ GeV in Fig.~\ref{fig:Anu_over_MR}.(1), and $560$ GeV in Fig.~\ref{fig:Anu_over_MR}.(2), respectively. The solid (red) curve represents Eq.~(\ref{eq:8}) and the dashed (green) curve represents Eq.~(\ref{eq:9}). It is seen from Figs.~\ref{fig:Anu_over_MR} that the constraint of CCB-1 is stronger than that CCB-2 
for $m_{\tilde{L}}=360$ GeV while the constraint of CCB-2 is stronger for $560$ GeV. The dependence of Eq.~(\ref{eq:8}) on
$m_{\tilde{\nu}_R}$ is trivial, and that of Eq.~(\ref{eq:9}) can be understood as follows. As we explained in Fig.~\ref{fig:gamma_ext}, 
$\gamma_{ext}$ becomes large as $m_{\tilde{L}}$ increases. Then, the right-hand side of Eq.~(\ref{eq:9}) increases due to a factor of
$1-\gamma^2$ in the denominator. This result is nontrivial, therefore we always have to check both constraints. 
The CCB-1 and CCB-2 constraint curves approach to $|A_\nu|/m_{\tilde{\nu}_R}=1$ as $m_{\tilde{\nu}_R}$ becomes large. In the large $m_{\tilde{\nu}_R}$ 
limit, the right-hand side of Eq.~(\ref{eq:8}) is dominated by $m_{\tilde{\nu}_R}$, and $\beta_{ext}$ goes to zero since $m_{\tilde{\nu}_R}$ 
appears in the denominator in Eq.~(\ref{eq:7}). 

\begin{figure}[t]
\begin{tabular}{c}
 \includegraphics[width=80mm]{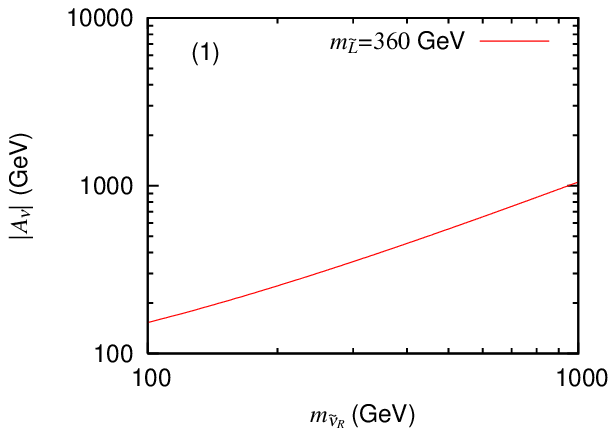} \\ 
 \includegraphics[width=80mm]{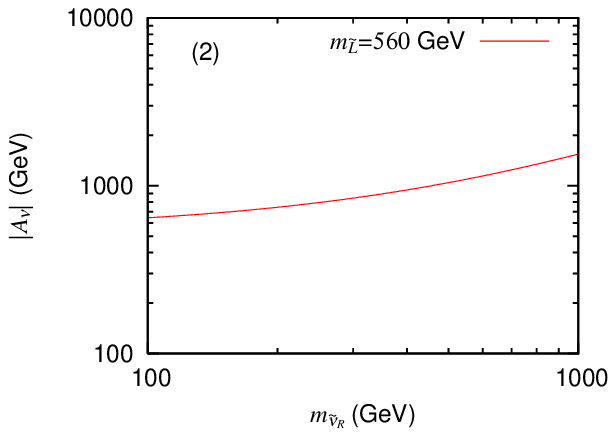} 
\end{tabular}
\caption{The upper bound on $A_\nu$ in terms of $m_{\tilde{\nu}_R}$. The left-handed slepton soft mass is $360$ GeV 
in Fig.~\ref{fig:Anu}.(1), and $560$ GeV in Fig.~\ref{fig:Anu}.(2), respectively.}
\label{fig:Anu}
\end{figure}
Figures \ref{fig:Anu} show the upper bound on $|A_\nu|$ in terms of $m_{\tilde{\nu}_R}$. The value of the left-handed slepton mass 
is indicated in the figures. The upper bound is more strict as $m_{\tilde{\nu}_R}$ is smaller and $m_{\tilde{L}}$ is smaller. In our 
example, $|A_\nu|$ must be smaller than $153$ GeV for $m_{\tilde{\nu}_R}=100$ GeV and $1550$ GeV for $m_{\tilde{\nu}_R} = 1000$ GeV in 
Fig.~\ref{fig:Anu}.(1) and $640$ GeV and $2020$ GeV in Fig.~\ref{fig:Anu}.(2), respectively. From numerical analysis, if the mass of 
the right-handed slepton is between a several $100$ GeV, $A_\nu$ term must be smaller than $1$ TeV.

\section{Summary and discussion}\label{sec:summary-discussion}
We have considered the $\nu$SSM where either Dirac or Majorana (s)neutrinos are introduced to the MSSM, and analyzed its scalar potential 
along the MSSM UFB/CCB directions as well as new CCB directions which appear due to non-vanishing vev's of right-handed sneutrinos. 

We have found that the MSSM UFB directions disappear and turn to CCB directions because the quartic term proportional to the square of the 
neutrino Yukawa coupling lifts up the potential for large values of fields. We have shown that depth of the minima along these directions is  
inversely proportional to the square of the neutrino Yukawa coupling, therefore it would be much deeper than that of EWSB. We derived 
necessary conditions to avoid the CCB minima along the MSSM UFB directions. The conditions impose constraints among the soft SUSY
breaking masses of the Higgses and the left-handed sneutrinos.

Then we have analyzed the potential along which the right-handed sneutrinos have non-vanishing vev's. We showed that CCB and incorrect 
EWSB minima exist along these directions. The minima are inversely proportional to the square of the neutrino Yukawa coupling hence 
very deep in cases of the Dirac neutrinos and Majorana neutrinos in TeV scale seesaw. Necessary conditions to evade these minima are 
derived for both Dirac and Majorana neutrinos. The conditions constrain the trilinear coupling of the sneutrino with respect to the soft 
masses. In the Majorana neutrino case, we found that one UFB direction appears due to the presence of the soft SUSY breaking mass terms of 
sneutrinos. A necessary condition to avoid the potential unbounded from below was also found.

In section~\ref{sec:numerical-analysis}, we have performed numerical analysis of the conditions to demonstrate a strategy to avoid 
the UFB and CCB minima. 
The strategy is that for a given set of the parameters consistent with the EWSB, firstly we check the condition from the MSSM UFB 
directions, (\ref{eq:21}) and (\ref{eq:22}). Next, we calculate $\gamma_{ext}$ using Eq.~(\ref{eq:11}) for the CCB-2 direction. 
Finally we check the conditions from CCB minima, (\ref{eq:8}) and (\ref{eq:9}). 
In figures \ref{fig:Anu_over_MR}, we have shown that the condition (\ref{eq:8}) is severer for $m_{\tilde{L}} = 360$ GeV 
and (\ref{eq:9}) is for $m_{\tilde{L}} = 560$ GeV.  
We have also shown in Figs.~\ref{fig:Anu} that the trilinear coupling is strictly constrained for smaller sneutrino masses. 
In the case that the right-handed sneutrinos are the lightest SUSY particles, this constraint is important to calculate their lifetime.

The conditions we found in this article are necessary conditions but not sufficient conditions. With these conditions satisfied, one can 
avoid dangerous UFB and CCB directions 
when radiative corrections are small compared with tree-level potential. As we mentioned in the introduction, it would be needed to 
include radiative corrections to obtain viable conditions at the electroweak scale. Since finite temperature effects would lift up 
the potential, it would be also important to consider finite temperature effects. 
We leave these for our future work.

\acknowledgments
The authors would like to thank Y.~Kanehata and Y.~Konishi for fruitful discussion and careful reading of this manuscript. 
T.~K. is supported in part by the Grant-in-Aid for Scientific Research No.~20540266 and the Grant-in-Aid for the Global COE 
Program "The Next Generation of Physics, Spun from Universality and Emergence" from the Ministry of Education, Culture, Sports, 
Science and Technology of Japan. T.~S is the Yukawa Fellow and the work of T.~S is partially supported by Yukawa Memorial Foundation.

\appendix
\section{Scalar potential of the MSSM}\label{apd:scal-potent-mssm}
In this appendix, we give notations of the scalars and the full scalar potential of the MSSM. 
The down-type and the up-type Higgs scalars are denoted as 
\begin{align}
 H_1 =
  \begin{pmatrix}
   H^1_1 \\
   H^2_1
  \end{pmatrix},\quad 
 H_2 = 
  \begin{pmatrix}
   H^1_2 \\
   H^2_2
  \end{pmatrix},
\end{align}
where $H_1^1$ and $H_2^2$ are electrically neutral. The left-handed squarks and the right-handed squarks are  
denoted as
\begin{align}
 \tilde{Q} =
  \begin{pmatrix}
   \tilde{u}_L \\
   \tilde{d}_L
  \end{pmatrix},\quad
 \tilde{u}_R,~\tilde{d}_R,
\end{align}
and the left-handed sleptons and the right-handed sleptons are denoted as 
\begin{align}
 \tilde{L} =
  \begin{pmatrix}
   \tilde{\nu}_L \\
   \tilde{e}_L
  \end{pmatrix},\quad
 \tilde{e}_R.
\end{align}

The scalar potential is divided into three parts which consist of 
$F$ terms, $D$ terms and soft SUSY breaking terms,
\begin{align}
 V = V_F + V_D + V_{soft}.
\end{align}
The $F$ term potential, $V_F$, is given by a sum of absolute square of all matter auxiliary fields,
\begin{align}
 V_F &= \sum_{i=\mathrm{matter}} |F_i|^2,
\end{align}
where
\begin{subequations}
\begin{align}
 F_{H^1_1}^\ast & = \mu H^2_2 + Y_e \tilde{e}_L \tilde{e}_R^\ast + Y_d \tilde{d}_L \tilde{d}_R^\ast ,\\
 F_{H^2_1}^\ast & = -\mu H^1_2 - Y_e \tilde{\nu}_L \tilde{e}_R^\ast - Y_d \tilde{u}_L \tilde{d}_R^\ast,\\
 F_{H^1_2}^\ast & = -\mu H^2_1 + Y_u \tilde{d}_L \tilde{u}_R^\ast,\\
 F_{H^2_2}^\ast & = \mu H^1_1 - Y_u \tilde{u}_L \tilde{u}_R^\ast,\\
 F_{\tilde{e}_R}   & = Y_e (H^1_1 \tilde{e}_L - H^2_1 \tilde{\nu}_L),\\
 F_{\tilde{e}_L}^\ast   & = Y_e H^1_1 \tilde{e}_R^\ast,\\
 F_{\tilde{\nu}_L}^\ast & = -Y_e H^2_1 \tilde{e}_R^\ast,\\
 F_{\tilde{d}_R}   & = Y_d (H^1_1 \tilde{d}_L - H^2_1 \tilde{u}_L),\\
 F_{\tilde{d}_L}^\ast   & = Y_d H^1_1 \tilde{d}_R^\ast + Y_u H^1_2 \tilde{u}_R^\ast,\\
 F_{\tilde{u}_R}      & = Y_u (H^1_2 \tilde{d}_L - H^2_2 \tilde{u}_L),\\
 F_{\tilde{u}_L}^\ast & = -Y_d H^2_1 \tilde{d}_R^\ast - Y_u H^2_2 \tilde{u}_R^\ast.
\end{align}
\end{subequations}
Here $\mu$ is a supersymmetric Higgs mass and $Y_i~(i=u,d,e)$ are Yukawa couplings.

The $D$ term potential, $V_D$, is given by a sum of square of all gauge auxiliary fields,
\begin{align}
 V_{D} &= \frac{1}{2}\bigg( (D^a_{SU(3)})^2 + (D^a_{SU(2)})^2 + (D_{U(1)})^2 \bigg),
\end{align}
where $a$ runs from $1$ to $8~(3)$ for $SU(3)~(SU(2))$ and the summation should be understood. The auxiliary fields, 
$D^a_{SU(3)}$, $D^a_{SU(2)}$ and $D_{U(1)}$, are given by
\begin{subequations}
\begin{align}
 D^a_{SU(3)} &= g_3 \left( \tilde{Q}^\dagger \frac{\lambda^a}{2} \tilde{Q} 
                          - \tilde{u}^\ast_R \frac{\lambda^a}{2} \tilde{u}_R
                          - \tilde{d}^\ast_R \frac{\lambda^a}{2} \tilde{d}_R \right), \\
 D^a_{SU(2)} &= g_2 \big( \tilde{Q}^\dagger T^a \tilde{Q} + \tilde{L}^\dagger T^a \tilde{L}
                + H_1^\dagger T^a H_1 + H_2^\dagger T^a H_2 \big),\\
 D_{U(1)} &= g_1 \left( \frac{1}{6} \tilde{Q}^\dagger \tilde{Q} -\frac{2}{3} \tilde{u}_R^\ast \tilde{u}_R 
           + \frac{1}{3} \tilde{d}^\ast_R \tilde{d}_R -\frac{1}{2}\tilde{L}^\dagger \tilde{L}\right. \nonumber \\ 
           &\qquad \left. + \tilde{e}_R^\ast \tilde{e}_R -\frac{1}{2}H_1^\dagger H_1 + \frac{1}{2}H_2^\dagger H_2 \right),
\end{align}
\end{subequations}
where $g_i~(i=1,2,3)$ is a gauge coupling constant, and $\lambda^a$ and $T^a$ are Gell-Mann and Pauli matrix respectively. 

The soft SUSY breaking terms, $V_{soft}$, are 
\begin{align}
  V_{soft} & = m^2_{H_1} H_1^\dagger H_1 + m^2_{H_2} H_2^\dagger H_2 
             + \big(B\mu H_1 \cdot H_2 + h.c.\big) \nonumber \\
          &~~+ m^2_{\tilde{Q}} \tilde{Q}^\dagger \tilde{Q} + m^2_{\tilde{u}_R} \tilde{u}_R^\ast \tilde{u}_R
             + m^2_{\tilde{d}_R} \tilde{d}_R^\ast \tilde{d}_R \nonumber \\
          &~~+ m^2_{\tilde{L}} \tilde{L}^\dagger \tilde{L} 
             + m^2_{\tilde{e}_R} \tilde{e}_R^\ast \tilde{e}_R \nonumber \\
          &~~+ \big(A_d Y_d H_1 \cdot \tilde{Q} \tilde{d}_R^\ast + A_u Y_u H_2 \cdot \tilde{Q} \tilde{u}_R^\ast \nonumber \\ 
          &~~+ A_e Y_e H_1 \cdot \tilde{L} \tilde{e}_R^\ast + h.c.\big),
\end{align}
where $m_i~(i=H_1,H_2,Q,u,d,L,E)$ are soft masses and $B\mu$ is a soft term for Higgses. A symbol ``dot'' represents an inner product 
for $SU(2)$ doublets, $A \cdot B = A^1B^2 - A^2 B^1$. The trilinear terms, $A_i~(i=u,d,e)$, are defined to be proportional 
to the corresponding Yukawa coupling. 

\section{Electroweak Symmetry Breaking of the MSSM}\label{sec:higgs-potent-electr}
We review the Higgs potential and the constraint from EWSB of the MSSM. 

The Higgs potential of the MSSM is given by
\begin{align}
  V &= m_1^2 H_1^2 + m_2^2 H_2^2 - (m_3^2 H_1 H_2 + h.c.) \nonumber \\
    &\quad + \frac{1}{8}(g_1^2 + g_2^2) \big( |H_1|^2 - |H_2|^2 \big)^2, \label{eq:17}
\end{align}
where 
\begin{subequations}
\begin{align}
 m_1^2 &= m_{H_1}^2 + |\mu|^2,\\
 m_2^2 &= m_{H_2}^2 + |\mu|^2,\\
 m_3^2 &= - B \mu.
\end{align}
\end{subequations}
A UFB direction is found along $D$ flat direction, namely,
\begin{align}
 |H_1|^2 = |H_2|^2,
\end{align}
and the potential becomes
\begin{align}
  V = (m_1^2 + m_2^2 - 2 |m_3^2|) |H_1|^2.
\end{align}
The potential is unbounded from below if the quadratic term is negative. Thus the constraint from UFB direction is given as 
\begin{align}
  m_1^2 + m_2^2 -2 |m_3^2| \geq 0.\label{eq:13}
\end{align}
This is so-called UFB-1 condition in \cite{Casas:1995pd}. For the EWSB to occur correctly, the potential must be a saddle point 
at the origin. The condition for such a saddle point is 
\begin{align}
  &\left( \frac{\partial^2 V}{\partial |H_1| \partial |H_2| }\right)^2 
  - \frac{\partial^2 V}{\partial |H_1|^2} \frac{\partial^2 V}{\partial |H_2|^2} \nonumber \\
 &=|m_3^2|^2 - m_1^2 m_2^2 > 0. \label{eq:14}
\end{align}
The EWSB vacuum is found by minimizing the potential, (\ref{eq:17}), with respect to the Higgses under the conditions, 
Eq.~(\ref{eq:13}) and (\ref{eq:14}). 

The EW symmetry is successfully broken at $|H_1| = v \cos\beta/\sqrt{2}$ and $|H_2| = v \sin\beta/\sqrt{2}$ if the 
following relations are satisfied,
\begin{align}
 m_1^2 + m_2^2 &= - \frac{2 m_3^2}{\sin 2\beta}, \label{eq:15} \\
 m_1^2 - m_2^2 &= - \cos 2\beta (m_Z^2 + m_1^2 + m_2^2), \label{eq:16} 
\end{align}
where $m_Z^2 = \frac{1}{4}(g_1^2 + g_2^2) v^2$. The minimum of the potential is obtained using Eq.~(\ref{eq:15}) and (\ref{eq:16}),
\begin{align}
  V_{\mathrm{real~min}} = - \frac{1}{2} \frac{m_Z^4}{g_1^2 + g_2^2} \cos^2 2\beta. \label{eq:18}
\end{align}

\section{scalar potential of the $\nu$SSM}\label{apx:scal-potent-numssm}
We list up here modifications of the MSSM scalar potential to the $\nu$SSM for Dirac neutrino and Majorana neutrino cases. 
The gauge auxiliary fields are the same as that of the MSSM because the right-handed (s)neutrinos are gauge singlets.

Firstly, we list up modifications in the Dirac neutrino case. Three of the matter auxiliary fields are replaced with
\begin{subequations}
\begin{align}
 F_{H^1_2}^\ast & = -\mu H^2_1 + Y_\nu \tilde{e}_L \tilde{\nu}_R^\ast + Y_u \tilde{d}_L \tilde{u}_R^\ast,\\
 F_{H^2_2}^\ast & = \mu H^1_1 - Y_\nu \tilde{\nu}_L \tilde{\nu}_R^\ast - Y_u \tilde{u}_L \tilde{u}_R^\ast,\\
 F_{\tilde{\nu}_L}^\ast & = -Y_e H^2_1 \tilde{e}_R^\ast - Y_\nu H^2_2 \tilde{\nu}_R^\ast,\\
 F_{\tilde{e}_L}^\ast   & = Y_e H^1_1 \tilde{e}_R^\ast + Y_\nu H^1_2 \tilde{\nu}_R^\ast,
\end{align}
\end{subequations}
where $\tilde{\nu}_R$ is the right-handed sneutrinos and $Y_\nu$ is the Yukawa couplings of neutrinos. The auxiliary fields of 
the right-handed neutrinos are added,
\begin{align}
 F_{\tilde{\nu}_R} = Y_\nu (H^1_2 \tilde{e}_L - H^2_2 \tilde{\nu}_L).
\end{align}
For the soft SUSY breaking term, a trilinear term $A_\nu$ and a soft mass $m_{\tilde{\nu}_R}$ for sneutrinos are added,
\begin{align}
  m^2_{\tilde{\nu}_R} \tilde{\nu}_R^\ast \tilde{\nu}_R + \big( A_\nu Y_\nu \tilde{L} \cdot H_2 \tilde{\nu}_R^\ast + h.c.\big).
\end{align}
Here $m_{\tilde{\nu}_R}$ is a soft SUSY breaking mass of the right-handed sneutrinos.

Next, we show two modifications in the Majorana neutrino case. One is on $F_{\tilde{\nu}_R}$ such as
\begin{align}
 F_{\tilde{\nu}_R}=Y_\nu (H^1_2 \tilde{e}_L - H^2_2 \tilde{\nu}_L) + M_R \tilde{\nu}^\ast_R,
\end{align}
where $M_R$ is the masses of the right-handed neutrinos. The other one is on the soft SUSY breaking term, that is, the following 
soft SUSY breaking mass is added,
\begin{align}
 \frac{1}{2}B_\nu M_R \tilde{\nu}^\ast_R \tilde{\nu}_R^\ast + h.c,
\end{align} 
where $B_\nu M_R$ is a soft mass for the right-handed sneutrinos.

\bibliographystyle{apsrev}
\bibliography{biblio}

\end{document}